\documentclass[pdflatex,sn-aps,iicol]{sn-jnl}

\usepackage{slashed}
\usepackage{graphicx}%
\usepackage{multirow}%
\usepackage{amsmath,amssymb,amsfonts}%
\usepackage{amsthm}%
\usepackage{mathrsfs}%
\usepackage[title]{appendix}%
\usepackage[svgnames]{xcolor}%
\usepackage{textcomp}%
\usepackage{manyfoot}%
\usepackage{booktabs}%
\usepackage{listings}%

\begin{document}

\title[Article Title]{Finite-energy sum rules at finite chemical potential and zero temperature}

\author[1,2,3]{\fnm{Alfredo} \sur{Raya}}

\author*[3,4]{\fnm{Cristian} \sur{Villavicencio}}\email{cvillavicencio@ubiobio.cl}

\affil[1]{\orgdiv{Instituto de F\'{i}sica y Matem\'aticas}, \orgname{Universidad
Michoacana de San Nicol\'as de Hidalgo}, \orgaddress{\city{Morelia}, \state{Michoac\'an} \postcode{58040}, \country{M\'{e}xico}}}

\affil[2]{\orgdiv{Facultad de Ingeniería Eléctrica}, \orgname{Universidad
Michoacana de San Nicol\'as de Hidalgo}, \orgaddress{\city{Morelia}, \state{Michoac\'an} \postcode{58040}, \country{M\'{e}xico}}}

\affil[3]{\orgdiv{Centro de Ciencias Exactas}, \orgname{Universidad del B\'io-B\'io}, \orgaddress{\postcode{Casilla 447}, \city{Chill\'an}, \country{Chile}}}

\affil[2]{\orgdiv{Departamento de Ciencias B\'asicas}, \orgname{Universidad del B\'io-B\'io}, \orgaddress{\postcode{Casilla 447}, \city{Chill\'an}, \country{Chile}}}

\abstract{In this article we explore the effect of 
%quark {\color{blue}(or baryon) }
chemical potential at zero temperature in the implementation  of {\em in-medium} effects in  the perturbative sector in finite energy sum rules.
For this purpose, we explore the axial, axial-pseudoscalar and pseudoscalar current correlators  involving charged pions.
The inclusion of non-normal ordered condensates with chemical potential effects in the operator mixing is considered.
%as well as in the scattering term.
As a result, the contribution of the operator mixing with chemical potential dependence cancels all the explicit chemical potential contribution of the perturbative sector, aligned with the so-called ``silver blaze problem".
We find an abrupt  transition when $\mu=\sqrt{s_0}/2$, with $s_0$ representing the hadronic continuum threshold.
Exploring beyond this critical chemical potential we found similarities with low-energy effective meson models at high chemical potential.
%In our findings, we consider the full mass dependence in the isospin symmetric mass approximation without radiative corrections.
}




\maketitle

% \documentclass[twocolumn,epj]{svjour}

% \usepackage{color}
% \usepackage{amssymb}
% \usepackage{amsmath}
% \usepackage{graphics}
% \usepackage{epsfig}
% \usepackage{bm}
% \usepackage{xcolor}
% \usepackage{slashed}
% \usepackage[colorlinks,citecolor=blue,linkcolor=blue,urlcolor=blue]{hyperref}

% \newcommand{\mums}{\mu_\textrm{\tiny MS}}

% \begin{document}

% \title{QCD sum rules at finite quark chemical potential and zero temperature}

% \author{Alfredo Raya$^{1,2}$ Cristián Villavicencio$^{2,3}$ and }

% \institute{ }

% \maketitle

After 50 years of the establishment of quantum chromodynamics (QCD) as the gauge theory of strong interactions~\cite{PhysRevLett.30.1343,PhysRevLett.30.1346}, the field of hadron phenomenology still possesses challenges and open questions that lead to the development of different techniques to address several issues, in particular, the nonperturbative sector.
Lattice simulations, functional methods and effective models each offer advantages and disadvantages in their implementation.
The study of the nonperturbative sector of the theory at finite baryon density using sum rules (SR) techniques has long data, where in the seminal papers~\cite{Cohen:1991js,Cohen:1991nk,Furnstahl:1992pi,Jin:1992id,Jin:1993up,Jin:1994bh}, the effects of baryon density are introduced in the condensates emerging from the operator product expansion (OPE) formalism, allowing to reach values near nuclear saturation density.  \cite{Cohen:1991js,Cohen:1991nk,Furnstahl:1992pi,Jin:1992id,Jin:1993up,Jin:1994bh,Kim:2001xu,Kim_2003,Thomas_2007,Mallik_2009,Zschocke_2011,Jeong:2012pa,Ohtani_2016,Cai_2019,Dominguez:2023bjb,Dominguez:2023tmt}.
In order to parameterize the baryon density to a baryon or quark chemical potential, in the SR framework, one can find several works with both, temperature and chemical potential. effects~\cite{Huang:1994vd,Mallik:2001gv,Zschocke:2002mn,Mallik:2004qj,Hilger:2010zb,Ayala:2011vs,Ayala:2013vra,Ayala:2016vnt,Bozkir:2022lyk}.
To the best of our knowledge, there are no works at zero temperature and finite chemical potential, so this scenario could reveal important features of in-medium sum rules.

In general, according with separation of short and long distance scales, or equivalently high and low energy scales, the perturbative sector carries on high-momentum information including medium effects, while condensates carry on the low energy information.
Wilson coefficients, which come together with condensates, involve high-energy information and these are obtained diagrammatically with pertutbative techniques.
The effects of the medium on the OPE contribution are manifested in condensate modifications, while the Wilson coefficients remain the same as in vacuum.
%\st{there is no reason a priori not to include explicit medium effects in the perturbative QCD sector.
%In the nonperturbative sector, however, the}
%Wilson coefficients suffer from infrared divergencies at finite temperature if one tries to abruptly take the chiral limit in the light-quark sector \cite{Adami:1990sv,Hansson:2021slq}, %\st{and that has been the main reason}
%reinforcing to consider medium effects only through the expectation value of the operators. %\st{in the SR literature.}

There are other considerations when focusing on the light-quark sector.
% For instance, when medium effects are present, the appearance of a scattering term with the medium in the  perturbative sector (which is dispersive) is of outmost relevance at finite temperature \cite{Bochkarev:1985ex,Dominguez:1989bz,Furnstahl:1989ji}.
% In the case of zero temperature but finite chemical potential, this term is also present.
%
% Also, 
The use of non-normal ordered condensates, instead of normal ordered condensates, produces an operator mixing contributing to the Wilson coefficients as well as in the perturbative sector \cite{Generalis:1990iy,Jamin_1993,GROZIN_1995,Narison:2002woh,Zschocke_2011,Hilger_2012,Gubler:2015qok}.
In principle, the operator mixing contribution to the perturbative sector can be explicitly medium-dependent, and this fact will be the main characteristic to be explored in this work.

Indeed, in this article we study the effects of quark or baryon chemical potential at zero temperature by studying the axial-vector and pseudoscalar correlators which involve the charged pions %\st{and $a_1$ mesons}
considering different scenarios in the perturbative sector, the nonperturbative sector and also considering operator mixing with explicit chemical potential contributions (OM($\mu$)), exploring specifically the role of the in-medium operator mixing compared with considering in-vacuum operator mixing (OM(0)) only.
%and the scattering term.
For this analysis, we employ the finite energy sum rules (FESR) without radiative corrections up to dimension 4 operators in the OPE sector.
%We consider also explicit mass dependence so that, in principle, our findings can be used in heavy quarks channels.
We have organized the this article as follows: in Sect.\,\ref{sec:invac} we revisit the formalism of FESR in vacum to set-up the notation and conventions. Modifications from a medium are discussed in Sect.\,\ref{sec:inmed}.
We specialize in a medium characterized by a finite quark chemical potential at zero temperature in Sect.\,\ref{sec:finmu}.
%An analysis of the results will be presented in Sect.\,\ref{sec:m2/s0->0}.
Conclusions are presented in Sect.\,\ref{sec:con}.
% followed by several appendices that complement the discussion and show details of calculations.

\section{In-vacuum formalism}\label{sec:invac}

Sum rules in QCD defined in the complex squared energy or $s$-plane, namely FESR, allow to relate hadron physics observables with the perturbative sector of QCD through the Cauchy residue theorem in that plane.
The FESR formalism considers the integration of the form factor along the closed contour dubbed as the {\it pac-man} contour in Ref.\,\cite{Dominguez:2018zzi,Dominguez:2018njv,Villavicencio:2020fcz,Villavicencio:2022gbr}, which consists in a circle in the mentioned plane excluding the positive real axis.
FESR connect the QCD and hadronic sector assuming the {\it duality principle}, where in the complex circle lies the QCD sector and in the discontinuity of the positive real axis lies the hadronic sector.
Then, by Cauchy residue theorem, the FESR read
\begin{align}
        \int_0^{s_0}\frac{ds}{\pi} \,s^{N} \text{Im}\,\Pi^\text{\tiny had}(s+i\epsilon)
    &=-\oint_{s_0}\frac{ds}{2\pi i}\,s^{N} \Pi^\text{\tiny QCD}(s)
    \label{eq:FESR_pac-man}
\end{align}
where $s^N$ is a weight function referred to as the kernel.

%FESR framework exhibit several advantages as compared to other QCD sum rules (QCDSR) approaches.
In vacuum, without  radiative corrections, FESR consistently truncates the OPE series, in the quark mass expansion, to every order corresponding to the power of $N$ in the weight function described in Eq.~\eqref{eq:FESR_pac-man} \cite{Ayala:2016vnt,Dominguez:2018zzi,Villavicencio:2020fcz}.
However, such a nice feature is broken when medium effects are taken into account.
Nevertheless, the set of SR in vacuum fixes the dimension of condensates that is relevant to consider.
Maybe a more useful feature, in computational terms, is that the order of integration of the loop internal momenta and the boundary integral can be switched \cite{Dominguez:2018njv,Villavicencio:2020fcz,Dominguez:2020sdf}.
This helps to avoid the calculation of the full form factor, which can be a complicated task in some circumstances.
One can then calculate easily the full set of FESR without the need of an additional  mass expansion, as commonly advocated when light quarks are considered.

\subsection{Non-normal ordered condensates and operator mixing}

The goal of SR is to establish relations among the hadronic and QCD form factors.
For the hadronic sector, hadron form factors are generically obtained from the spectral function as
\begin{equation}
    \Pi^\text{\tiny{had}}(p) = \int_{0}^\infty ds\, \frac{\rho(s)}{s-p^2},
    \label{eq:contour}
\end{equation}
with the spectral function defined for each resonance as
\begin{equation}
    \rho(s)=\frac{1}{\pi}\text{Im}\,\Pi(s).
\end{equation}
On the other hand, for the QCD sector, the correlators are split into the perturbative QCD (pQCD) contribution and the nonperturbative sector described by OPE.
Recall that the idea behind OPE is to parameterize the nonperturbative sector in series of operators as
\begin{align}
\Pi(p^2) &= \Pi_\text{\tiny pQCD}(p^2,\nu) +\sum_{n>0}  C_n(p^2,\nu)\,\langle :\! {\cal O}_n\!:\rangle
\end{align}
where $\nu$ above is a scale factor (identified with the $\overline{\text{MS}}$ scale in our discussion), being the $\Pi(p)$ the Fourier transformation of the current-current correlator $\Pi(x,0)$. 
Vacuum is considered immersed in a gluon background  and condensates though in principle are non-local, are expanded around $x=0$.

Unlike the case with heavy quarks, considering the expectation value of the normal-ordered quark condensate for light quarks presents several challenges due to their small current masses. 

Long-distance effects are expected to be encapsulated in the OPE condensates, while short-distance dynamics are captured by perturbative QCD and the Wilson coefficients.
Therefore, to achieve a proper separation of scales, the parameters relevant to the short-distance regime should be consistent with that scale; however, terms proportional to $\ln m_q$ correspond to long-distance contributions.

Non-analytic behavior in mass functions and infrared-divergent terms emerge in the chiral limit. Moreover, for two light-quark flavors, certain contributions become ill-defined in this limit~\cite{Narison:2002woh} because the outcome depends on the order in which the quark masses approach zero; taking the limit $m_u \to 0$ before $m_d \to 0$ gives different results than taking the simultaneous limit $m_u = m_d \to 0$.

To address all the aforementioned issues, one considers {\it non-normal ordered} condensates in place of the standard normal-ordered ones
\cite{Generalis:1990iy,Jamin_1993,GROZIN_1995,Zschocke_2011,Hilger_2012,Gubler:2015qok}.
Specifically, in the case of quark bilinear operators, the non-normal-ordered quark condensate is defined as
\begin{equation}
    \langle \bar q \,\Gamma q\rangle \equiv \langle \Omega|: \bar q\,\Gamma q:|\Omega\rangle
    +\langle \Omega|{\cal T}\, \bar q\,\Gamma q|\Omega\rangle,
    \label{eq:NNoc_qq}
\end{equation}
where $|\Omega\rangle$ stands for the gluon-saturated vacuum and $\Gamma$ denotes a general operator.
The last term represents the full quark propagator in the gluon background, traced with the operator $\Gamma$. 
Detailed expressions for the operator mixing are provided in Sec.\,\ref{sec:finmu}

The non-normal ordered quark condensate eliminates the light-quark mass logarithms and replaces them with the $\overline{\text{MS}}$ prescription, making the condensates scale-dependent. Consequently, all singularities and the ill-defined behavior in the chiral limit are resolved.

The full propagator in \eqref{eq:NNoc_qq} is expanded in powers of the background gluon fields, resulting in an operator mixing of the form
\begin{equation}
    \langle : {\cal O}_n:\rangle = \langle  {\cal O}_n(\nu)\rangle
    +c_{n0}(\nu)
    +\sum_{m>0} c_{nm}(\nu)\, \langle {\cal O}_m(\nu)\rangle ,
    \label{eq:NOC(NNOC)}
\end{equation}
such that $c_{n0}$ contributes to the pQCD sector.
The correlation function now reads
\begin{align}
       \Pi(p^2)     &= \tilde \Pi_\text{\tiny pQCD}(p^2,\nu) +\sum_{n>0} \tilde  C_n(p^2,\nu)\,\langle {\cal O}_n  (\nu)\rangle,
\label{eq:OPE(p,mu)}
\end{align}
where the non-normal ordered condensates are scale-dependent and renormalizable.

\subsection{Axial-vector and pseudoscalar correlators}

To setup the SR, let us define the combined time-ordered correlators in momentum space as
\begin{align}
    \Pi_{\mu\nu}^{A}(p)
    &= i\int d^4x\, e^{ip\cdot x}\langle {\cal T} A_\mu(x) A^\dag_\nu(0) \rangle , \\
    \Pi_{5\nu}(p)
    &= i\int d^4x\, e^{ip\cdot x}\langle {\cal T} \,i\partial^\mu A_\mu(x) \,A_\nu^\dag (0) \rangle , \\
       \Psi_{5}(p)
    &= i\int d^4x\, e^{ip\cdot x}\langle {\cal T} \,i\partial^\mu A_\mu(x)\, [i\partial^\nu A_\nu]^\dag (0) \rangle ,
\end{align}
with the charged axial-vector current defined as $A_\mu\equiv A_\mu^{1}-iA_\mu^2$.
The light-quark correlators in the axial-vector and pseudoscalar channels involve the pions as the lowest-energy participants.
These particles, in different models, participate in the pseudoscalar as well as in the axial-vector channels.
This double militancy, as a consequence of the partial conservation of the axial current, provides more information in terms of the number of SR to be dealt with.
As usual, inserting a complete set of intermediate on-shell hadronic states between currents, we consider only a single pion state.
The  matrix elements for the currents involved are
\begin{align}
    \langle 0|A_\mu(x)|\pi^+(p)\rangle & =\sqrt{2} f_\pi ip_\mu e^{-ip\cdot x}, \\
     \langle 0|\partial^\mu A_\mu(x)|\pi^+(p)\rangle & =\sqrt{2} f_\pi m_\pi^2 e^{-ip\cdot x},
\end{align}
and in the same way $A^\dag$ has non-vanighing matrix element with a $\pi^-$ state.
Alternatively, the positively charged axial-vector current can be expressed, like in chiral perturbation theory, in terms of hadronic fields.
For positive charged pions,
\begin{align}
    A_\mu &=-f_\pi\partial_\mu \pi^+ ,\\
  \partial^\mu A_\mu &=\sqrt{2}f_\pi m_\pi^2 \pi^+.
\end{align}
On the other hand, the positively charged axial-vector current and the divergence of the axial-vector current for QCD degrees of freedom are, respectively,
\begin{align}
A_\mu   &= \bar d\gamma_\mu\gamma_5  u,\\
 \partial^\mu A_\mu &= (m_u+m_d)\bar d i\gamma_5 u.
\end{align}

Explicitly, for the hadron sector, correlators are
\begin{align}
    \Pi_{\mu\nu}^{A}(p)
    &= -p_\mu p_\mu\frac{2f_\pi^2}{p^2-m_\pi^2},\\
    \Pi_{5\nu}(p)
    &= -p_\nu \frac{2f_\pi^2m_\pi^2}{p^2-m_\pi^2},\\
       \Psi_{5}(p)
   &= - \frac{2f_\pi^2m_\pi^4}{p^2-m_\pi^2},
\end{align}
%with $M_{a_1}^2\equiv m_{a_1}^2-i\sqrt{p^2}\,\Gamma_{a_1}(p^2)$ and $M_\pi^2=m_\pi^2-i\sqrt{p^2}\,\Gamma_\pi$.
%Hereafter we consider $\Gamma_\pi\to 0$, unless we want to explore medium effects on pion decay width.
Notice that the axial-vector pseudoscalar correlator is factorized as $\Pi_{5\nu}= p_\nu \Pi_5$.
The axial-vector correlator, in turn, can be factorized in terms of its transverse and diagonal parts.
Under this factorization, the diagonal part is related with the pseudoscalar correlators $\Pi_5$ and $\Psi_5$ through Ward identities.
Thus, all the physically relevant information is contained in the axial-vector correlator. 
However, it is more feasible to factorize it as a longitudinal and diagonal components
\begin{equation}
    \Pi_{\mu\nu}^A =p_\mu p_\nu \Pi_0 + g_{\mu\nu}\Pi_1,
\end{equation}
and work directly with the form factors $\Pi_0$, $\Pi_5$ and $\Psi_5$.

The non-vanishing set of FESR which consider upto dimension 4 operators are $\Pi_0$ with $N=0$ and $N=1$, $\Pi_5$ with $N=0$ and $\Psi_5$ with $N=0$, corresponding to
\begin{align}
    2f_\pi^2 & = \frac{s_A}{4\pi^2},
        \label{eq:F(0)Pi_0-vac}\\
   2f_\pi^2m_\pi^2 & = \frac{s_A^2}{8\pi^2}-2m\langle {\bar qq}\rangle -\frac{1}{12}\langle G^2\rangle,\\
    2f_\pi^2 m_\pi^2 &= 2m\left\{\frac{3m s_P}{4\pi^2}-2\langle {\bar qq}\rangle \right\},
        \label{eq:F(0)Pi_5-vac}\\
    2f_\pi^2 m_\pi^4 &= 4m^2\left\{\frac{3s_P^2}{16\pi^2}-m\langle {\bar qq}\rangle +\frac{1}{8}\langle G^2\rangle\right\},
        \label{eq:F(0)Psi_5-vac}
\end{align}
where the hadronic threshold is different for the axial-vector correlator, denoted as $s_A$ and for the others involving the pseudoscalar correlator, denoted as $s_P$.
We will refer hereafter $s_0$ to the hadronic threshold in general, which can be $s_A$ or $s_P$.
We neglected corrections $m_q^2/s_0$.

The average quark condensate and the average quark mass are defined as
\begin{equation}
    \langle {\bar qq}\rangle =\frac{1}{2}(\langle \bar uu\rangle + \langle\bar dd\rangle),
    \quad m \equiv \frac{1}{2}(m_u+m_d),
\end{equation}
respectively.
The gluon condensate is defined as
\begin{equation}
    \langle G^2\rangle = \left\langle \frac{\alpha_s}{\pi}G_{\mu\nu}^a G^{a\,\mu\nu}\right\rangle
    =\frac{2\alpha_s}{\pi} \left(\langle B^2\rangle -\langle E^2\rangle\right),
\end{equation}
$B_i^a$ and $E_i^a$ denoting the components of the chromomagnetic and chromoelectric fields, respectively.

Notice that Eq.~\eqref{eq:F(0)Pi_5-vac} is the Gell-Mann--Oakes--Renner relation, being the term $\sim m^2 s_P$  negligible.

In what follows, we consider $m_u\approx m_d$ for simplicity in the final results, but as a general rule, we keep $m_d\neq m_u$ along the calculation.
Although $m_d/m_u\sim 2$, the terms involving the quark mass difference are of the form $(m_d-m_u)(\langle\bar dd\rangle - \langle\bar uu \rangle)$ and the rest of contributions with an explicit mass difference are suppressed as $m_q^2/s_0$.
Therefore, the isospin symmetric approximation is reasonable in these channels.

\subsection{The continuum hadronic threshold}
\label{subsec:s0}

The continuum hadronic threshold $s_0$ is the main phenomenological parameter in FESR.
At finite temperature, it behaves as an order parameter for the confinement-deconfinement transition \cite{Ayala:2016vnt,Dominguez:2018zzi}.
This parameter is defined as the energy threshold where there is an overlap of resonances.
This definition, however, is fuzzy sometimes because the exact point at which resonances overlap can be actually a wide region. Therefore, it is commonly obtained from the FESR as one of the unknown parameters.
This fact makes the present formalism appealing when in-medium effects are taken into account because the hadronic threshold becomes medium dependent and, as mentioned above, it serves as an indicator of confinement or deconfinement, depending on its behavior when the appropriate parameters vary.

Since contour integration is performed around the hadronic threshold, it is customary to choose the QCD subtraction scale as $\nu=\sqrt{s_0}$.
However, here we are dealing with two hadronic thresholds, so the appropriate subtraction scale in this case is to take the average
\begin{equation}
    \nu=\frac{\sqrt{s_A}}{2}+\frac{\sqrt{s_P}}{2}\sim 1\text{ GeV}.
    \label{eq:nu_vac}
\end{equation}
If we do not consider radiative corrections, with pion mass $m_\pi=0.14$\,GeV and pion decay constant $f_\pi=0.092$\,GeV as inputs, the axial hadronic threshold $\sqrt{s_A}=0.82$\,GeV  which can be directly obtained from equation \eqref{eq:F(0)Pi_0-vac}.
If we consider running quark mass at the scale in \eqref{eq:nu_vac}, the hadronic threshold in the pseudoscalar channel is $\sqrt{s_P}=1.24$\,GeV, and consequently, the subtraction scale is $\nu=1.03$\,GeV.

It is important to ensure that the hadronic thresholds do not exceed the next resonance pole mass, in our case, the $a_1(1260)$ in the axial channel and $\pi(1300)$ in the pseudoscalar channel.

\section{In-medium formalism}~\label{sec:inmed}

\begin{figure}
    \centering
    \includegraphics[scale=0.5]{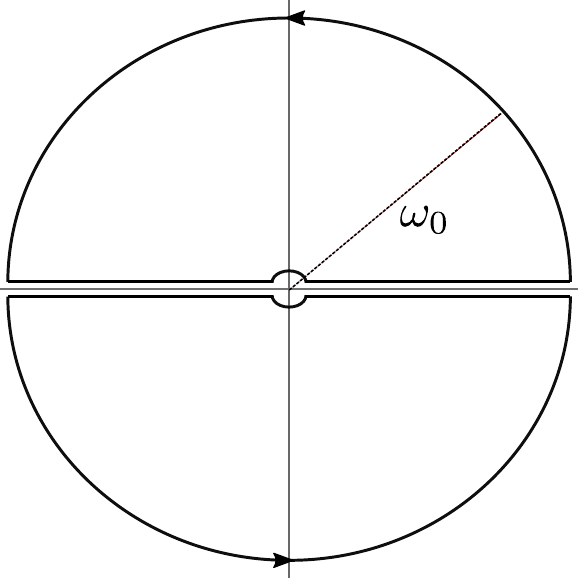}
    \caption{The integration contour in the complex $p_0$-plane. The large circle has radius $w_0$, The small circle avoids the origin.}
    \label{fig-contour}
\end{figure}

When taking into account a heat bath at finite temperature and/or chemical potential, Lorentz symmetry is broken. As a consequence,
 the form factors depend on the 4-momentum as $\Pi(p_0,\boldsymbol{p}^ 2)$.
A commonly used strategy in SR is to separately consider the even and odd components in $p_0$ and use the same {\it pac-man} contour for each components with $\boldsymbol{p}=0$ and $s=p_0^2$ \cite{Cohen:1991js,Cohen:1991nk,Furnstahl:1992pi,Jin:1992id,Jin:1993up,Jin:1994bh}.
Alternatively, in this article we adopt the integration in the complex $p_0$-plane, as it is shown in Fig.~\ref{fig-contour}, instead on $p^2$ complex plane \cite{Furnstahl:1992pi}.
Integration along this path yields completely equivalent results to the usual contour, but presents certain advantages when calculating the contributions of the QCD sector. 
In particular, it helps to perform contour integration  before the internal loop momentum integration, that turns out very convenient.
The FESR using this contour are
\begin{equation}
\begin{split}
    {\int\hspace{-10pt}-}_{\!\!\!-\omega_0}^{\,\omega_0} \,\frac{d\omega}{\pi} \,\omega^{n+1} \text{Im}\,\Pi^\text{\tiny had}(\omega+i\epsilon,\boldsymbol{p})
    \qquad\qquad\\
    =-\oint_{\omega_0}\frac{d\omega}{2\pi i}\,\omega^{n+1} \Pi^\text{\tiny QCD}(\omega,\boldsymbol{p})\\
    +\underset{\omega=0}{\text{Res}}
    \left[\omega^{n+1}\Pi^\text{\tiny QCD}(\omega,\boldsymbol{p})\right],
    \label{eq:FESR}
\end{split}
\end{equation}
where the symbol ${\int\hspace{-9pt}-}$ denotes the principal value excluding $\omega=0$.
The radius of the large circle is $\omega_0 = \sqrt{s_0+\boldsymbol{p}^2}$.
Usually, the above relation \eqref{eq:FESR} is calculated in the frame $\boldsymbol{p}=0$.
In vacuum, one obtain the same results of Eq.~\eqref{eq:FESR_pac-man} with $N=2n$.

We emphasize the importance ofthe pole in $\omega=0$.
Although it has no impact in vacuum, it has important consequences at finite temperature, giving rise to the term representing the scattering with the thermal bath of quarks with space-like momentum known as {\it the scattering term} \cite{Bochkarev:1985ex,Dominguez:1989bz,Furnstahl:1989ji}.
This term is also present at finite chemical potential, and thus can be interpreted as scattering of quarks with the dense bath.
% For more details see Appendix\,\ref{app.Integrals}.

\subsection{Hadronic sector}
\label{subsec:had_correlators}

In accordance with the contour considered for in-medium correlators in terms of $p_0$, the hadron form factor is now defined as
\begin{equation}
    \Pi^\text{\tiny had}(\omega,\boldsymbol{p}^2)=\int_0^\infty dp_0^2\,\frac{\rho(p_0,\boldsymbol{p}^2)}{p_0^2-\omega^2}.
\end{equation}
% Under such definition, we have
% \begin{equation}
%     \text{Im}\,\Pi_i^\text{\tiny had}(p_0+i\epsilon,\boldsymbol{p}^2) =\text{sign}(p_0)\theta(p^2)\text{Im}\,\Pi_i(p^2) .
% \end{equation}

% \subsection{Axial-vector and pseudoscalar correlators}

In order to setup the FESR framework including a medium, we start from the correlators in momentum space for the hadron sector.
We consider an effective in medium pion model which preserves symmetry breaking effects which is the most often used.
This is based on self-energy corrections and field strength renormalization.
% {\color{blue}
% It is important to remark that for our main purpose of this work, which is the study of the role of ON($\mu$), the results are
% independent of which hadronic model is used since we are focusing in the QCD sector.
% Hadronic model will be relevant for a posteriori annalysis
% }

Chemical potential breaks Lorentz symmetry, splitting Lorentz structures in time and space.
Therefore, the axial-vector current splits into temporal and spatial components parametrized with different temporal and spatial pion decay constants
$f_t$ and $f_s$.
The decay constant and the particle velocities are related to the temporal and spatial decay constants as
\begin{equation}
    f_\pi=\sqrt{f_s f_t},\qquad v_\pi=\sqrt{f_s/f_t},
\end{equation}
being the first relation a definition for the mutual decay constant and the second relation appears as a consequence due to Ward identity \cite{Pisarski:1996mt,Son:2002ci,Lee:2003rj}.
With all these considerations, the in-medium form factors have the structure
% \sp{which are given as}
\begin{align}
    \Pi_{\mu\nu}^{A}(p)
    &=
    % -2f_{a_1}^2 \frac{P'_\mu   P'_\nu -M_{a_1}^2  g'_{\mu\nu}}{p_0^2-v_{a_1}^2\boldsymbol{p}^2-M_{a_1}^2}
    % \nonumber \\ &\quad
            - 2f_\pi^2 v_\pi^{-2}\frac{P_\mu P_\mu}{p_0^2-v_{\pi}^2\boldsymbol{p}^2-m_\pi^2}, \\
            % &\nonumber\\
    \Pi_{5\nu}(p)
    &=  -2f_\pi^2 v_\pi^{-2}\frac{m_\pi^2 \, P_\nu}{p_0^2-v_{\pi}^2\boldsymbol{p}^2-m_\pi^2},\\
       % &\nonumber\\
       \Psi_{5}(p)
   &= - 2f_\pi^2 v_\pi^{-2}\frac{m_\pi^4}{p_0^2-v_{\pi}^2\boldsymbol{p}^2-m_\pi^2},
\end{align}
%{\color{green}
%with
%\begin{align}
%    P'_\mu &=\omega_\mu + v_{a_1}^2\bar p_\mu,\\
%    P_\mu &=\omega_\mu + v_{\pi}^2\bar p_\mu,\\
%     g'_{\mu\nu} & = u_\mu u_\nu +v_{a_1}^2\bar g_{\mu\nu},
%\end{align}
%}
with $P_0=p_0$ and $P_i =v_{\pi}^2p_i$
%\st{and  $\bar g_{\mu\nu}\equiv g_{\mu\nu}-u_\mu u_\nu$ are the spatial components of the Minkowski metric tensor.
%With these definition, the contraction of $\Pi^A_{\mu\nu}$ and $\Pi_{5\nu}$ with the momentum $p$ removes the pole in the $a_1$ form factor and preserves the Ward identity of the pion form factor.}

%
Next, we can separate the axial correlator and mixed axial-pseudoscalar correlator in independent components.
We just separate temporal and spatial components.
Then, $\Pi^A$ and $\Pi_5$ can be factorized as
\begin{align}
    \Pi^A_{\mu\nu} &= (\omega_\mu \bar p_\nu+\omega_\nu \bar p_\mu)\Pi_0^{ts} + \bar p_\mu \bar p_\nu \Pi_0^{ss}
    \nonumber \\ &\qquad \qquad\qquad
    +u_\mu u_\nu \Pi_{1}^t + \bar g_{\mu\nu}\Pi_1^s,\\
    &\nonumber\\
    \Pi_{5\nu} &= \omega_\nu \Pi_5^t+\bar p_\nu\Pi_5^s,
    \label{eq:in-medium_factorization}
\end{align}
where $u_\mu\equiv g_{\mu 0}$ is the four-velocity in the rest frame,  $\omega_\mu \equiv p_0 u_\mu$ is the time component of the momentum, $\bar p_\mu \equiv p_\mu -\omega_\mu$ is the spatial part of the momentum and  $\bar g_{\mu\nu}\equiv g_{\mu\nu}-u_\mu u_\nu$ are the spatial components of the Minkowski metric tensor.

Like in the vacuum case,  two of the $\Pi^A$ components are related with $\Pi_{5}$ Because of the Ward identity. 
So, we only consider $\Pi_0^{ts}$ and $\Pi_0^{ss}$.
Notice that the structure $\omega_\mu\omega_\nu$ is the same as $u_\mu u_\nu$. Therefore, there is no $\Pi_0^{tt}$ because it is the same structure of $\Pi_1^t$.

\subsection{QCD sector}

% {\color{red}The condensates and the perturbative part,  as well as the coefficients in Eq.~\eqref{eq:NOC(NNOC)} that contribute to the perturbative part, should become now medium dependent.}
The condensates and the perturbative part become now medium-dependent.
This means that, if we denote as $\Xi$ the set of parameters that defines the external medium, like temperature and chemical potentials, the
 correlator is then %\st{approximated as}
\begin{equation}
    \Pi(p,\Xi) = \tilde\Pi_\text{\tiny pQCD}(p,\nu,\Xi) +\sum_{n>0} \tilde C_n(p,\nu)\,\langle {\cal O}_n(\nu,\Xi)\rangle .
    \label{eq:OPE_in-medium}
\end{equation}
% where now we explicitly remark that it depends on $p$ instead of $p^2$ because of Lorentz symmetry breaking due to the medium.
Thus, the Wilson coefficients do not depend on medium parameters, but the condensate does.
The question here is wether the coefficients obtained from the operator mixing in Eq.\,\eqref{eq:NOC(NNOC)} are medium dependent or not, in particular the contribution to pQCD part $c_{n0}$.
We address this question in the next section.

In the case of OPE, in addition to the usual quark and gluon condensates, thermal and/or density effects generate new condensates, i.e.,
\begin{equation}
\langle u^\dag u\rangle,\quad
\langle d^\dag d\rangle, \quad
\langle \Theta_{00}^q\rangle , \quad
\langle \Theta_{00}^g\rangle,
\end{equation}
where the first two condensates are the quark-number densites.
The traceless quark energy-momentum tensor is defined as
\begin{equation}
    \Theta_{\mu\nu}^q = \sum_{q=u,d}\bar q\left[\frac{1}{2}\left(i\gamma_\mu D_\nu + i\gamma_\nu  D_\mu\right)
    -\frac{g_{\mu\nu}}{d}m_q \right]q.
\end{equation}
For gluons, its traceless energy-momentum tensor is
\begin{equation}
    \Theta_{\mu\nu}^g= \frac{g_{\mu\nu}}{d} G^a_{\alpha\beta} G^{a\,\alpha\beta}
    -G^a_{\mu\beta} G_\nu^{a\,\beta},
\end{equation}
whereas the gluon energy in terms of chromoelectric and chromomagnetic field reads
\begin{align}
   \langle \Theta_{00}^g\rangle \,= \frac{1}{d}\left[ 2\langle B^2\rangle +(d-2)\langle E^2\rangle\right].
\end{align}
For short-hand notation, we define the energy contributions of the quarks and gluons as
\begin{align}
    \langle\theta_q \rangle &\equiv \langle\Theta_{00}^q\rangle, \\
    \langle \theta_g \rangle &\equiv \frac{\alpha_s}{\pi}\langle \Theta_{00}^g\rangle,
\end{align}
respectively.

With these ingredients, we are ready to set up our framework in the forthcoming section.

\section{Finite-energy sum rules at finite chemical potential}\label{sec:finmu}

We are now interested in the specific details of the FESR framework with a finite chemical potential. For
baryons, the chemical potential enters directly in the QCD sector through the quark propagators.
Pions are unaffected directly unless the $u$ and $d$ chemical potentials are different, which could happen indirectly through the modification of their parameters.
The time-ordered quark propagator at finite chemical potential \cite{Shuryak:1980tp,PhysRevD.42.2881,Mizher:2013kza,Mizher:2018dtf,Villavicencio:2023jxj} is
\begin{align}
    S_q(p) &=\frac{i(\slashed{p}+m_q)}{p^2-m^2+i\epsilon p_0(p_0-\mu_q)}
    \label{eq:Sq(mu)1}\\
        &=(\slashed{p}+m_q)\bigg[\frac{i}{p^2-m_q^2+i\epsilon}
        \nonumber\\& \qquad-2\pi\theta(p_0(\mu_q-p_0))\delta(p^2-m_q^2)\bigg].
        \label{eq:Sq(mu)2}
\end{align}
Alternatively, one can calculate loop corrections at finite temperature, perform the sums over Matsubara frequencies and take the limit $T\to 0$.
Both approaches provide the same results.

For what comes next, we consider the same chemical potential for the $u$ and $d$ quarks and express all results in terms of
\begin{align}
    \mu\equiv\frac{1}{2}(\mu_u+\mu_d)=\frac{1}{3}\mu_B,
\end{align}
where $\mu_B$ is the baryonic chemical potential.
Hereafter we assume that $\mu>0$ but all the results are independent of the sign of $\mu$.
The details of the calculation of the QCD sector are described in Appendix\,\ref{app.Integrals}.

% Looking at Eq.\,\eqref{eq:one-loop_final}, the final expression for the contour and internal momentum integrals of the generic one-loop contribution, it can be seen that the final result is different if the chemical potential is larger or smaller than a certain critical value defined as
% \begin{equation}
%     \mu_c = \sqrt{s_0}/2.
% \end{equation}
Below we analyze what is this critical chemical potential.

Next, we explore the effect of finite chemical potential from the operator mixing.
This contributions comes from considering non-normal ordered quark condensates.
The operator mixing in this work is generated by the following operator
\begin{align}
    &\langle:\bar q \,\gamma_{\mu}iD_\nu \,q:\rangle
    =\langle\bar q \,\gamma_{\mu}iD_\nu \,q\rangle\nonumber \\
    &\qquad +\text{tr}\int\frac{d^dk}{(2\pi)^d}\,\gamma_\mu i\tilde D_\nu \,S_q(k,\nu;G),
    \label{eq:O0}
\end{align}
where $S_q(p,\nu;G)$ is the full quark propagator in a gluon background, $D$ and $\tilde D$ are the covariant derivatives in configuration and momentum spaces, respectively, and the trace is performed over spin and color indexes.
%
%The quark operator that contributes to the operator mixing is described in Eq,\,\eqref{eq:O0}.
Expanding in the background gluon operator, it can be written as
\begin{align}
    &\langle :\bar q \gamma_\mu iD_\nu q :\rangle = \langle \bar q \gamma_\mu iD_\nu q \rangle
    + a_{\mu\nu}(\mu) \nonumber \\
   &\qquad + b_{\mu\nu}(\mu) \langle G^2 \rangle +c_{\mu\nu}(\mu)\langle \theta_g\rangle +\dots,
\end{align}
with
\begin{align}
    a_{\mu\nu}(\mu)   &   =   \frac{N_c m^4}{8\pi^2}\left[\frac{1}{d-4}+\ln\left(\frac{m}{\nu}\right)-\frac{3}{4}\right]g_{\mu\nu}
    \nonumber\\         & \quad   +\theta(\mu-m)\bigg\{\frac{N_c}{8\pi^2}\bigg[m^4\ln\left(\frac{\mu+p_F}{m}\right)\nonumber
    \\         & \quad  -\mu p_F(\mu^2+p_F^2)\bigg]g_{\mu\nu} +\frac{N_c}{3\pi^2}\mu p_F^3\,\bar g_{\mu\nu}\bigg\},
    \label{eq:a_mu-nu}
\end{align}
where $p_F\equiv \sqrt{\mu^2-m^2}$.
Here we only show the OM($\mu$) contribution to pQCD since the Wilson coefficients in OPE sector remain the same as in vacuum. Furthermore,
\begin{align}
    b_{\mu\nu}(0)       &=\frac{1}{48}\langle G^2\rangle g_{\mu\nu},\\
    &\nonumber\\
    c_{\mu\nu}(0)       &=\frac{1}{3}\langle\theta_g\rangle \left[
        \frac{1}{d-4}+\ln\left(\frac{m}{\nu}\right)\right]
        \nonumber\\
        &\qquad\times \left(u_\mu u_\nu - \frac{\bar g_{\mu\nu}}{d-1}\right).
\end{align}
More information of operator mixing can be found in \cite{Generalis:1990iy,Jamin_1993,GROZIN_1995,Zschocke_2011,Hilger_2012,Gubler:2015qok}.

To simplify the notation, we write the FESR for each form factor as
\begin{equation}
    F^{(n)}_\text{\tiny had}[\Pi] = F^{(n)}_\text{\tiny pQCD}[\Pi]+F^{(n)}_{qq}[\Pi]+F^{(n)}_{GG}[\Pi]+\dots
    \label{eq:FESR notation}
\end{equation}
described by Eq.\,(\ref{eq:FESR}), where $n$ denotes the power of the wheight function $\omega^{n+1}$.
The left-hand side stands for the hadronic sector and on the right-hind side, we have the contributions of the perturbatibe QCD sector, the quark condensates and gluon condensates.

Let us start by analyzing the contributions with odd $n$.
The non-vanishing contributions correspond to $F^{(1)}[\Pi^{ts}_0]$, $F^{(1)}[\Pi^{t}_5]$ and  $F^{(1)}[\Pi^{s}_5]$, yielding the following set of equations
\begin{align}
   0    &=\langle d^\dag d\rangle - \langle u^\dag u\rangle,
    \label{eq:Fqq-1-ts}\\
    0   &=  4m^2[\langle d^\dag d\rangle -\langle  u^\dag u\rangle],
      \label{eq:Fqq-1-t}\\
    0   &=  4m^2[\langle d^\dag d\rangle -\langle  u^\dag u\rangle].
    \label{eq:Fqq-1-s}
   % F^{(1)}_{qq}[\Psi_5]        &= 0
\end{align}
We can see from the left hand side of these equations, that hadronic FESR vanishes because we do not consider isospin chemical potential and therefore no isospin asymmetry.
In this case, the set of equations above indicate that the number of $u$-quarks is the same as the number of $d$-quarks.

Now, for even $n$, it is more convenient to express the results as a combination of temporal and spatial form factors, namely, sums and differences of the form
\begin{align}
\Pi_0^{\pm}&=\frac{1}{2}(\Pi_0^{ts}\pm\Pi_0^{ss}),\\
\Pi_5^\pm &= \frac{1}{2}(\Pi_5^{t}\pm\Pi_5^{s}).
\end{align}
The following equations are derived by the FESR for the average of temporal and spatial in $\Pi_0$ and $\Pi_5$, for $\Psi_5$ and for the difference between time and spatial form factor components of $\Pi_0$ and $\Pi_5$.
In the notation of FESR, this is $F^{(0)}[\Pi_0^{+}]$, $F^{(2)}[\Pi_0^{+}]$, $F^{(0)}[\Pi_5^{+}]$, $F^{(0)}[\Psi_5]$, $F^{(0)}[\Pi_0^{-}]$, $F^{(2)}[\Pi_0^{-}]$ and  $F^{(0)}[\Pi_5^{-}]$:
\begin{align}
    f_\pi^2(1+v_\pi^2)  &=  \frac{s_A}{4\pi^2}\,C_{\Pi_0^+}^{(0)},
                           \label{eq:FESR_mu_Pi0+0}
                        \\
    f_\pi^2(1+v_\pi^2)m_\pi^2  &=  \frac{s_A^2}{8\pi^2}\,C_{\Pi_0^+}^{(2)}
                                    -2m\langle\bar qq\rangle
                        \nonumber\\&\qquad
                                    -\frac{1}{12}\langle G^2\rangle
                                    +\frac{4}{9}\langle\theta_g\rangle,
                             \label{eq:FESR_mu_Pi0+2}   \\
    f_\pi^2(v_\pi^{-2}+1)m_\pi^2   &=  2m\left\{\frac{3m s_P}{4\pi^2}\,C_{\Pi_5^+}^{(0)}
                -2\langle \bar qq\rangle \right\},
                \label{eq:FESR_mu_Pi5+0}
\end{align}
\begin{align}
      2f_\pi^2v_\pi^{-2}m_\pi^4   &=   4m^2\bigg\{\frac{3\,s_P^2}{16\pi^2}\,C_{\Psi_5}^{(0)}
                                            -m\langle \bar qq\rangle
                                            -2\langle \theta_q\rangle
                                    \nonumber\\&
                                            +\frac{1}{8}\langle G^2\rangle
                                            +\frac{1}{6}\left[
                                                1-4\ln\left(\frac{s_P}{\nu^2}\right)\right]
                                                    \langle\theta_g\rangle\bigg\},
                                                    \label{eq:FESR_mu_Psi0}
\end{align}
\begin{align}
    f_\pi^2(1-v_\pi^2)  &=  \frac{s_A}{4\pi^2}\,C_{\Pi_0^-}^{(0)}
                             +\frac{4}{9s_A}\langle \theta_g\rangle,
                             \label{eq:Pi00-}
                        \\
    f_\pi^2(1-v_\pi^2)m_\pi^2  &=  \frac{s_A^2}{8\pi^2}\,C_{\Pi_0^-}^{(2)}
                                    -\frac{4}{3}\langle \theta_q \rangle
                                \nonumber   \label{eq:Pi02-}\\ &
                                    +\frac{1}{9}\left[\frac{11}{3}-4\ln\left(\frac{s_A}{\nu^2}\right)\right]\langle\theta_g\rangle,
                                \\
    f_\pi^2(v_\pi^{-2}-1)m_\pi^2   &= %\frac{3m^2\,s_P}{2\pi^2}\,C_{\Pi_5^-}^{(0)}
                                                \frac{8m^2}{3s_P}\langle \theta_g\rangle,
                                                \label{eq:Pi5-}
\end{align}
where the coefficients $C_\Pi^{(n)}$ are chemical potential dependent.
At zero chemical potential,  $\langle \theta_g\rangle, \langle\theta_q\rangle\to 0$ and   $v_\pi,C_\Pi^{(n)}\to 1$, so Eqs.\,\eqref{eq:FESR_mu_Pi0+0}-\eqref{eq:FESR_mu_Psi0} reduce to Eqs.\,\eqref{eq:F(0)Pi_0-vac}-\eqref{eq:F(0)Psi_5-vac}, except in the case of the difference of temporal and spatial, where $C_{\Pi_0^-}^{(n)}\to 0$ in-vacuum.
We left the leading order contribution in the QCD side in Eq.\,\eqref{eq:Pi5-} for further analysis, but keeping in mind that it is highly suppressed (at least at low chemical potential).

As we mentioned before, there will be an abrupt transition when the chemical potential exceeds certain value in each channel, which is related to the hadronic threshold.
The coefficients $C_\Pi^{(n)}$ change if $\mu<\mu_c$ or $\mu>\mu_c$  where
\begin{equation}
\mu_c=
 \begin{cases}
 \sqrt{s_A}/2,\qquad \text{axial-vector channel,}\\
 \sqrt{s_P}/2,\qquad \text{pseudoscalar channel.}
 \end{cases}
\end{equation}

Table\,\ref{table:CPi^n} shows the different values of the coefficients $C_\Pi^{(n)}$ for the cases considering OM($\mu$) as well as considering OM(0).
These cases are separated if the chemical potential is below or above the critical chemical potential in their respective channels.
These values are assumed to have their tabulated values for $\mu>m$, otherwise, all the coefficients become 1, except the $\Pi_0^-$ ones which vanish.
In the same Table, the approximation $m^2/s_0\sim 0$ is considered, and consequently if $\mu>\mu_c$ then $m/\mu\sim 0$.

\begin{table}
    \centering
\begin{tabular}{|c||c |c||c| c|}
\hline
\multicolumn{1}{|c||}{ $C_{\Pi}^{(n)}$}&\multicolumn{2}{c||}{in-medium OM}&\multicolumn{2}{|c|}{in-vacuum OM} \\
 &   $\mu<\mu_c$ &   $\mu>\mu_c$  &   $\mu<\mu_c$ &   $\mu>\mu_c$\\
\hline &&&&\\
$C_{\Pi_0^+}^{(0)}$ &  $1$ & $\frac{7}{10}\left(\frac{\mu}{\mu_c}\right)^2$ & 1 &   $\frac{7}{10}\left(\frac{\mu}{\mu_c}\right)^2$   \\&&&&\\
$C_{\Pi_0^+}^{(2)}$ &  $1$  & 0  & 1 &   0    \\&&&&\\
$C_{\Pi_5^+}^{(0)}$ & 1 &   $\left(\frac{\mu}{\mu_c}\right)^2$ &  $1-\frac{\mu p_F}{\mu_c^2} $ & 0   \\&&&&\\
$C_{\Psi_5}^{(0)}$  & 1 &   $\left(\frac{\mu}{\mu_c}\right)^4$&  $1-\frac{\mu p_F^3}{\mu_c^4}$ &  0   \\&&&&\\
$C_{\Pi_0^-}^{(0)}$ &   $0$ &  $\frac{3}{10}\left(\frac{\mu}{\mu_c}\right)^2$  & 0 &   $\frac{3}{10}\left(\frac{\mu}{\mu_c}\right)^2$   \\&&&&\\
$C_{\Pi_0^-}^{(2)}$ &   0 &  $\left(\frac{\mu}{\mu_c}\right)^4$    &   $-\frac{\mu p_F^3}{\mu_c^4}$ & 0
\\&&&&\\
% $C_{\Pi_5^-}^{(0)}$ &   0 & 0  &   0 &   0   \\&&&&\\
\hline
\end{tabular}

\caption{pQCD coefficients considering OM($\mu$) (second and third columns) and considering OM(0) (fourth and fifth columns).
In all the cases $\mu>m$, and terms $\sim m^2/ s_0$ has been neglected.}
\label{table:CPi^n}
\end{table}

\subsection{Analysis for $\mu<\mu_c$}

Starting with the first case in Table\,\ref{table:CPi^n} which is considering pQCD both, OM($\mu$) and OM(0) (second and fourth column) we can observe that there is a remarkable fact:
For the chemical potential below the critical value, the coefficients $C_\Pi^{(n)}$ are exactly the same as in the vacuum case.
This is not only in the approximation $m^2/s_0\sim 0$ but also if we consider the general mass dependence, for which the full FESR expressions are shown in Appendix\,\ref{app:full_FESR}.

The fact that the explicit dependency of the chemical potential is canceled for $\mu<\mu_c$ is exactly what the Silver Blaze problem proposes: At zero temperature, any explicit dependence on the baryon chemical potential manifests only beyond certain large critical value \cite{COHEN_2005,Gunkel:2019enr}.
This may be considered as a good proof that OM($\mu$) must be taken into account, mainly because this result is independent of the hadronic model, but bearing in mind that we are demonstrating this for FESR only and it is not clear if other QCD sum rules approaches reproduce this cancellation of the chemical potential dependency.
To deepen into this scenario, let us  explore it in more detail.

The most significant equation in the set of FESR is Eq.\,\eqref{eq:Pi5-}, because its right-hand  side is basically negligible.
Since we are dealing with a general value of the chemical potential, including low values, the only possibility is that $v_\pi\sim 1$ unless pion mass or pion decay constant tend to vanish.
Combining with Eq.\,\eqref{eq:Pi00-}, this inmediately implies that $\langle \theta_g\rangle =0$.
Now, the difference arises from Eq.\,\eqref{eq:Pi02-}: If we consider OM($\mu$), it implies that $\langle \theta_q\rangle=0$.
On the other hand, if we consider OM(0), the result gives $\langle \theta_q\rangle = -\frac{3}{2\pi^2}\mu p_F^3$.
This is in contradiction with previous QCD sum rules approaches at finite baryon density which predict positive $\langle \theta_q\rangle$\cite{Cohen:1991js,Cohen:1991nk,Furnstahl:1992pi,Jin:1992id,Jin:1993up,Jin:1994bh,Kim:2001xu,Kim_2003,Thomas_2007,Mallik_2009,Jeong:2012pa,Ohtani_2016,Cai_2019,Dominguez:2023bjb,Dominguez:2023tmt}.

In the rest of FESR equations, the differences are subtle, but with the arguments given above, it is clear that not considering chemical potential dependence on the operator mixing is in contradiction with general well established results.
The fact that there is no explicit chemical  potential dependence for $\mu<\mu_c$ does not necessary mean that there is no implicit dependence in condensates and hadronic threshods.
Indeed, this is what usual sum rules at finite baryonic density establish.

\subsection{Analysis for $\mu>\mu_c$}

Once accepted the need of OM($\mu$) for $\mu<\mu_c$, we will assume same condition but considering now the case where $\mu>\mu_c$ (chemical potential bigger than critical value in both channels) in the chiral limit, but relating pion mass with quark mass as $m_\pi^2\sim m$ like in vacuum, or more concretely,  
\begin{equation}
m_\pi^2=  2m B_\pi.
\end{equation}
with $B_\pi$ finite in the chiral limit.

Although we do not know what happens with the hadronic thresholds in this phase,  the chiral limit assumes that $s_0\gg m^2$.
So, even if $s_0$ decreases considerably, the analysis assumes that it is much larger than $m^2$.

Other thing to be taken into account, although there is no incidence in this analysis, is the scale factor.
One can notice from Eq.\,\eqref{eq:nu_vac} that the subtraction scale is $\nu=\mu_c^\textrm{\tiny ($P$)}+\mu_c^\textrm{\tiny ($A$)}$.
Since now the chemical potential is larger than both critical values, the subtraction scale is then
\begin{equation}
 \nu =2\mu.
\end{equation}
This subtraction scale has been suggested in pQCD calculations at finite chemical potential \cite{Ipp:2003jy,Kurkela:2009gj}, and here this scale emerges naturally.

The procedure is to cancel $m$ wherever it can be canceled and then set $m\to 0$.
Starting once again with Eq.\,\eqref{eq:Pi5-}, diving by $2m$ and setting $m\to 0$, we get $f_\pi^2(v_\pi^{-2}-1)B_\pi =0$.
This equation imposes one of the following conditions: $f_\pi=0$, $B_\pi =0$ or $v_\pi=1$.
The decay constant cannot be zero because of Eq.\,\eqref{eq:FESR_mu_Pi0+0}.
If $v_\pi= 1$, the condensate $\langle \theta_g\rangle$ becomes negative, as can be seen in Eq.\,\eqref{eq:Pi00-}, in contradiction with the established values.
So, the only choice is $B_\pi=0$, which means that the pion mass vanishes, in concordance with chiral-symmetry restoration.
Under this assumption, and combining the set of equations, we obtain
\begin{align}
 f_\pi^2 &= \frac{\mu^2}{2\pi^2}(1+a),
 \label{eq:fpi_mu>muc}
 \\
 v_\pi^2 &= \frac{\frac{2}{5}-a}{1+a},
 \label{eq:vpi_mu>muc}
 \\
 \langle G^2\rangle &= \frac{16}{3}\langle \theta_g\rangle,
 \\
 \langle\bar qq\rangle &=0,
 \\
 \langle \theta_q\rangle &= \frac{3\mu^4}{2\pi^2}+
 \frac{1}{3}\left[\frac{11}{12}-\ln\left(\frac{s_A}{\nu^2}\right)\right]\langle\theta_g\rangle,
 \\
 0 &=\left[\frac{1}{3}-\ln\left(\frac{s_P}{s_A}\right)\right]\langle \theta_g\rangle,
 \label{eq:sP/sA}
\end{align}
with the parameter $a$ defined as
\begin{equation}
    a=\frac{4\pi^2}{9}\frac{\langle \theta_g\rangle}{s_A\,\mu^2}.
\end{equation}
If we assume that $\langle\theta_g\rangle \neq 0$, as it is expected, then the last equation provides the following relation among the hadronic thresholds
\begin{equation}
    s_P = s_A e^{1/3}.
\end{equation}
It is interesting that Eq.\,\eqref{eq:fpi_mu>muc} resembles results of effective low-energy QCD models at high $\mu$ \cite{Son:1999cm,Hong:1999ei,Manuel:2000wm,Beane:2000ms,Alford:2007xm}.
Although these models are constructed considering color-superconductivity, the numerical match is remarkable, where in those models $f_\pi \sim 0.2\mu$.
If we consider the usual restrictions for the pion velocity in Eq.\,\eqref{eq:vpi_mu>muc} $0<v_\pi^2<1$, then $0<a<\frac{2}{5}$.
In this range of values, $0.225 <f_\pi/\mu < 0.266$.
For the particular value of the pion velocity in the at the speed of sound in ultrarelativistic fluids $v_\pi^2=1/3$ we have $a=1/20$ and $f_\pi=0.231\mu$.

\bigskip
Let us see what happens if we assume $m_\pi^2 \sim m^2$, as it has been also predicted in some effective dense meson models \cite{Son:1999cm,Hong:1999ei,Huang:2004ik}.
% , which is sometimes proposed in effective dense meson models [..].
If we set $m_\pi^2=  4m^2 C_\pi$, we need to include the next contribution in the expansion $m^2/s_0$.
The only relevant contribution appears in Eq.~\eqref{eq:Pi5-}, which now becomes
$f_\pi^2(v_\pi^{-2}-1)C_\pi = 2\langle \theta_g\rangle/3s_P$.
Repeating the procedure but now replacing $m_\pi^2$ by $4m^2C_\pi$, canceling the masses $m$ wherever it can be canceled, and then setting $m\to 0$, we obtain the same relations of Eq. \eqref{eq:fpi_mu>muc}-\eqref{eq:sP/sA}.
The result for $C_\pi$ as a function of the parameter $a$  becomes
\begin{equation}
    C_\pi=e^{1/3}\frac{3a}{1+a}\frac{2-5a}{3+5a},
\end{equation}
which is smaller than 0.2 for the allowed values of the parameter $a$.

\section{Discussion and conclusions}
\label{sec:con}

In this article we have obtained the finite-energy sum rules at finite quark or baryon chemical potential at zero temperature for the axial-axial, axial-pseudoscalar and pseudoscalar-pseudoscalar correlators including OPE up to dimension 4 operators.
%
% Considering chemical potential dependent operator mixing of the OPE, the FESR integrated Wilson coefficient has an infrared divergence when $\mu=m$.
% This  {\color{red}reinforces the condition} that all the medium effects in OPE are contained in the operators only.
% %
% On the other hand,

In the perturbative sector, if one considers chemical potential dependence in the operator mixing generated from the normal ordered quark condensate, the contour integrated coefficients become independent of the chemical potential for values less than a critical value, which is $\mu_c=\sqrt{s_A}/2$ in the axial-vector channel and $\mu_c=\sqrt{s_P}/2$ in the pseudoscalar channel.

For $\mu>\mu_c$, the appearance of an explicit chemical potential dependence arises.
This is in accordance with the Silver Blaze problem, which claims that there is no explicit baryon chemical potential up to a certain critical value.
Although this problem is often addressed in terms of the free energy, this should not be considered a coincidence.

Both critical values, if we consider vacuum hadronic thresholds, correspond to critical baryon chemical potential of $\mu_B \approx 1.23$\,GeV in the axial-vector channel and $\mu_B\approx 1.86$\,GeV in the pseudoscalar channel, which are considerable large compared with  critical chemical potential for chiral restoration predicted by most of models, which turn to be near the nucleon mass.
Nevertheless, the high critical chemical potential presented here best closely approximates the transitions predicted with Schwinger-Dyson equations \cite{Bender_1998,Jiang:2013xwa,Gunkel:2019xnh,Gunkel:2020wcl}.
It is important to keep in mind that, even if there is no explicit chemical potential contribution, implicit contribution coming from condensates and from the hadronic thresholds, and consequently the hadronic parameters, are possible. 
Therefore, the hadronic thresholds may diminish with chemical potential, diminishing the values of the the critical chemical potentials. 
At finite temperature, the hadronic threshold behavior is related to the deconfinement phase transition; we have to be careful in the interpretation of this critical potential, but it is more likely that the critical value here is related to the deconfinement transition.
Another possibility is that the critical chemical potentials only indicate the validity threshold of the model.

% Notice that this also reinforces the  assumption adopted in QCDSR at finite baryon density and zero temperature literature that, in fact, the pQCD sector does not depend explicitly on the baryon density, {\color{red} at least for values lower than a certain critical scale}.
% On th other hand, works at finite temperature should be reviewed in order to incorporate in-medium contribution from the oprator mixing in the perturbative sector.

In the case of $\mu$ larger than both critical chemical potentials, our analysis in the chiral limit exhibits vanishing of the pion mass, as expected, but a linear dependency in the chemical potential for the pion decay constant.
This linear behavior coincides with chiral meson models under high chemical potential which predicts $f_\pi\sim 0.2\mu$ and vanishing pion mass like in this work.

The fact that chemical potential dependence in the operator mixing must be incorporated suggests that other in-medium problems should be revisited.
The appropriate combination of SR with other approaches that can provide the point-wise chemical potential evolution of the condensates becomes the natural next step to consider.
For the time being, the best candidate for our purposes is the Nambu--Jona-Lasinio model.
This is a work under consideration, and all findings will be reported elsewhere.

\bigskip

\subsection*{Acknowledgments}
    AR acknowledges  %Consejo Nacional de Humanidades, Ciencia y Tecnología (México) under grant CF-2023-G-433 as well as 
    Consejo de la Investigación Científica (UMSNH, México) under project 18371. C.V. Acknowledges financial support from ANID/FONDECYT under grant 1250206.

%%%%%%%%%%%%%%%%%%%%%%%%%%%%%%%%%%%%%%%%%%%%%%%%%%%
\begin{appendices}

\section{One-loop integrals}\label{app.Integrals}

% {\color{blue}In order to obtain the scattering term propperly, it is important that internal $u$ and $d$ quark internal energy in the pQCD diagrams differ, so any simplification like equal masses, and rest-frame must be done after FESR integration contour.
% In general, chiral limit is considered and rest frame is applied later.
% Here we consider different quark masses in rest frame and approximate equal or vanishing masses later.}

\begin{figure}
    \centering
    \includegraphics[scale=.5]{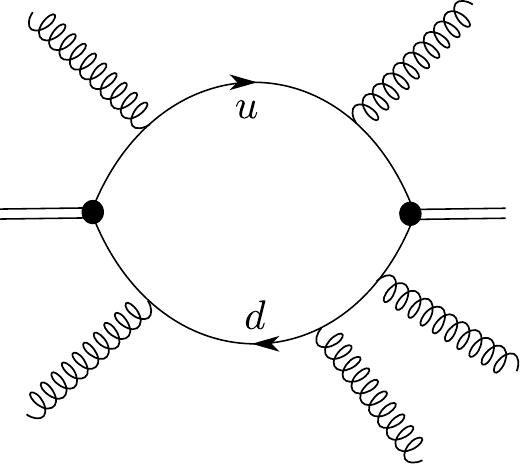}
    \caption{Generic one-loop diagram with gluon insertions.}
    \label{fig:1-loop}
\end{figure}

% Notice that 
There are diagrams with gluon insertions that appear with higher powers of the propagators as can be seen in the general diagram described in Fig.\,\ref{fig:1-loop}.
In this case, we can recast it as a simple one-loop diagram formed by two propagators considering derivatives of the squared quark masses.
Keeping aside quark masses in the numerator (or renaming them), a diagram with $a$ insertions of gluons in the $u$-quark line, and $b$ gluon insertions in the $d$-quark line can be written as
\begin{equation}
    \Pi=\frac{1}{(a-1)!(b-1)!}\left(\frac{\partial}{\partial m_u^2}\right)^a \left(\frac{\partial}{\partial m_d^2}\right)^b \tilde \Pi ,
\end{equation}
where in this case we use the propagators described in Eq.\,\eqref{eq:Sq(mu)1}.
Let us define a generalized one-loop contribution 
\begin{align}
    \tilde\Pi[f] = \int\frac{d^d k}{(2\pi)^d}f(k_0,p_0,\boldsymbol{k},\boldsymbol{p})S_u(k)S_d(k+p),
\end{align}
where the propagators  can be set as described in Eq.\,(\ref{eq:Sq(mu)2}).
The function $f$ consists of powers of internal and external energy and momentum.
Defining the chemical potential dependent part as $\Delta\tilde \Pi=\tilde\Pi[f]-\tilde\Pi[f]_{\mu=0}$, after integration in $k_0$ we get
\begin{strip}
\begin{align}
  \Delta\tilde \Pi=-\int\frac{d^3k}{(2\pi)^3} \Bigg[ &
  \frac{f(E_u,p_0,\boldsymbol{k},\boldsymbol{p})\theta(\mu-E_u)}{2E_u[(p_0+E_u)^2-E_d^2+i\epsilon]}
  % \nonumber\\ \quad
  +\frac{f(E_d-p_0,p_0,\boldsymbol{k},\boldsymbol{p})\theta(\mu-E_d)}{2E_d[(p_0-E_d)^2-E_u^2+i\epsilon]}
  \nonumber\\ &
  +\frac{ f(E_u,p_0,\boldsymbol{k},\boldsymbol{p})}{4E_u E_d}\theta(\mu-E_u)\theta(\mu-E_d)\,
  2\pi i\delta(p_0+E_u-E_d)\Bigg].
  \label{eq:int_k0}
\end{align}
Integrating the above equation along the contour of Fig.\,\ref{fig-contour} for the QCD sector we obtain
\begin{align}
    \oint_\Gamma \frac{d\omega}{2\pi i}\Delta\tilde\Pi
    &=\int\frac{d^3k}{(2\pi)^3} \frac{1}{4 E_u E_d}  \Big[ 
    \left\{ 
         f(E_u,-E_+,\boldsymbol{k},\boldsymbol{p}) \theta(\omega_0-E_+)
        -f(E_u,E_-,\boldsymbol{k},\boldsymbol{p}) \theta(\omega_0-|E_-|)
    \right\} \theta(\mu-E_u)
        \nonumber\\ &\qquad\qquad \qquad\quad
    -\left\{ 
        f(-E_u,E_+,\boldsymbol{k},\boldsymbol{p}) \theta(\omega_0-E_+) 
        -f(E_u,E_-,\boldsymbol{k},\boldsymbol{p}) \theta(\omega_0-|E_-|)
    \right\} \theta(\mu-E_d)         \Big],
\end{align}
where the integration path is $\Gamma=C(\omega_0)-C(\epsilon)$, where $C(\omega_0)$ is the big circle,  and $C(\epsilon)$ is the small circle in Fig.\,\ref{fig-contour} that produces the residue in $p_0=0$.
Notice that the delta function in the last line of Eq.\,\eqref{eq:int_k0} vanishes in this contour since it does not pass through the real axis.

\bigskip
Finally, integrating in the momentum $\boldsymbol{k}$ and taking the  limit $\boldsymbol{p}\to 0$,
\begin{align}
    \oint_\Gamma \frac{d\omega}{2\pi i}\Delta\tilde\Pi%_{\boldsymbol{p}\to 0}
    = & \left[\int_0^{\bar k_u}\frac{d|\boldsymbol{k}|\boldsymbol{k}^2}{E_uE_d} f(E_u,-E_+,|\boldsymbol{k}|,0)
        -\int_0^{k_u}\frac{d|\boldsymbol{k}|\boldsymbol{k}^2}{E_uE_d} f(E_u,E_-,|\boldsymbol{k}|,0)
    \right]\frac{\theta(\mu-m_u)}{8\pi^2}
    \nonumber\\&
    -\left[\int_0^{\bar k_d}\frac{d|\boldsymbol{k}|\boldsymbol{k}^2}{E_uE_d} f(-E_u,E_+,|\boldsymbol{k}|,0)
        -\int_0^{k_d}\frac{d|\boldsymbol{k}|\boldsymbol{k}^2}{E_uE_d} f(E_u,E_-,|\boldsymbol{k}|,0)
    \right]\frac{\theta(\mu-m_d)}{8\pi^2},
    \label{eq:one-loop_final}
\end{align}
\end{strip}
where the different energy terms are defined as
\begin{align}
    E_u &=\sqrt{\boldsymbol{k}^2+m_u^2},\\
    E_d &=\sqrt{(\boldsymbol{k}+\boldsymbol{p})^2+m_d^2},\\
    E_\pm &= E_d\pm E_u,
\end{align} 
and with the integration limits defined as 
\begin{align}
    k_q &= \sqrt{\mu^2-m_q^2},\\
    k_s &=\sqrt{\frac{s_0}{4}-\frac{m_d^2+m_u^2}{4}+\frac{(m_d^2-m_u^2)^2}{4s_0}},\\
    \bar k_q &= \text{min}(k_q,k_s).
\end{align}
The integral involving $\bar k_q$ can be expressed as
\begin{align}
    \int_0^{\bar k_q} \ldots  = &\quad \theta\left(\mu-\frac{\sqrt{s_0}}{2}\pm \frac{m_d^2-m_u^2}{2\sqrt{s_0}}\right)\int_0^{k_s}\ldots
    \nonumber\\ &
    +\theta\left(\frac{\sqrt{s_0}}{2}\pm \frac{m_d^2-m_u^2}{2\sqrt{s_0}}-\mu\right)\int_0^{k_q}\ldots
\end{align}
where in the Heavyside functions ``+'' is for $q=d$ and ``$-$'' is for $q=u.$
Here we can identify for $m_d^2-m_u^2\ll s_0$ that the condition is for $\mu$ larger or smaller than $\mu_c$.

% The integration path is $\Gamma=C(\omega_0)-C(\epsilon)$, where $C(\omega_0)$ is the big circle,  and $C(\epsilon)$ is the small circle in Fig.\,\ref{fig-contour} that produces the residue in $p_0=0$.

% Notice that the delta function in the last line of Eq.\,\eqref{eq:int_k0} vanishes in this contour since it does not pass through the real axis.
% It can be expressed as $2\pi i \delta(x)=\frac{1}{x+i\epsilon}-\frac{1}{x-i\epsilon}$, integrate in the contour and then set $\epsilon\to 0$. 
% {\color{red}This produces no effect since in the function $f$, the possible poles in $p_0$ were removed in  contour. 

It is worth mentioning again that the procedure with Matsubara frequencies at finite temperature provides the same results when taking the zero temperature limit.

\section{Full FESR expressions}
\label{app:full_FESR}

The full FESR are calculated for the QCD sector considering equal mass approximation up to dimension 4 in the OPE.
The notation used is specified in Eq.\,\eqref{eq:FESR notation}.
Perturbative QCD is considered for chemical potential including OM($\mu$).

\subsection{pQCD}

FESR contribution in the pQCD sector for $\mu<\mu_c$ are
\begin{align}
   & F^{(0)}_\text{\tiny pQCD}[\Pi_0]    =  \frac{s_A}{4\pi^2}\;\sigma_A^3 ,   \\
    &\nonumber\\
    & F^{(2)}_\text{\tiny pQCD}[\Pi_0]     =  \frac{s_A^2}{8\pi^2}\;\frac{\sigma_A}{2}(3-\sigma_A^2)
         \nonumber\\ & \qquad
            +\frac{3m^4}{4\pi^2}\left[3- 4
   \ln\left(\frac{\sqrt{s_A}}{\nu}\frac{1+\sigma_A}{2}\right)\right],\\
    &\nonumber\\
     &F^{(0)}_\text{\tiny pQCD}[\Pi_5]     =    \frac{3m^2 s_P}{2\pi^2}\;\sigma_P
     \nonumber \\ & \qquad +\frac{3m^4}{\pi^2}\left[
     1-2
   \ln\left(\frac{\sqrt{s_P}}{\nu}\frac{1+\sigma_P}{2}\right)
   \right],\\
   &\nonumber\\
  & F^{(0)}_\text{\tiny pQCD}[\Psi_5]     = \frac{3m^2s_P^2}{4\pi^2}\;\frac{\sigma_P}{2}(1+\sigma_P^2)
    \nonumber\\ & \qquad
         +\frac{3m^6}{2\pi^2}\left[3-4
   \ln\left(\frac{\sqrt{s_P}}{\nu}\frac{1+\sigma_P}{2}\right)\right],
\end{align}
where we have introduced the notation
\begin{equation}
    \sigma_A\equiv \sqrt{1-\frac{4m^2}{s_A}},
    \quad \text{and}\quad
    \sigma_P\equiv \sqrt{1-\frac{4m^2}{s_P}}.
\end{equation}
Here, we have suppressed the temporal and spatial indexes $t$ and $s$ because the results are the same.

The FESR contributions for the pQCD sector for $\mu~>~\mu_c$ are 
\begin{align}
    F^{(0)}_\text{\tiny pQCD}[\Pi_0^{ts}]    &=\frac{1}{\pi^2}\frac{p_F^3}{\mu},\\
    &\nonumber\\
    F^{(0)}_\text{\tiny pQCD}[\Pi_0^{ss}]    &= \frac{2}{5\pi^2}\frac{p_F^5}{\mu^3},\label{eq:FpQCD^0[Piss]-mu>muc}
    \\ & \nonumber\\
% \end{align}
% \begin{align}
      F^{(2)}_\text{\tiny pQCD}[\Pi_0^{ts}]  &  =\frac{\mu p_F}{\pi^2}\left(2\mu^2+m^2\right)
        \nonumber\\ & \!\!\!
        +\frac{3m^4}{4\pi^2}\left[3-4\ln\left(\frac{\mu+p_F}{\nu}\right)\right],\\
    % F^{(2)}_\text{\tiny pQCD}[\Pi_0^{ss}]    &=\frac{\mu p_F}{\pi^2}\left(2\mu^2+m^2\right)
    %     \nonumber\\ &\quad
    %     +\frac{3m^4}{4\pi^2}\left[3-4\ln\left(\frac{\mu+p_F}{\Lambda}\right)\right]
    &\nonumber\\
    F^{(2)}_\text{\tiny pQCD}[\Pi_0^{ss}]    &=\frac{\mu p_F}{\pi^2}\left(-2\mu^2+5m^2\right)
        \nonumber\\ &\!\!\!
        +\frac{3m^4}{4\pi^2}\left[3-4\ln\left(\frac{\mu+p_F}{\nu}\right)\right],\\
    % F^{(2)}_\text{\tiny pQCD}[\Pi_0^{ss}]    &=\frac{\mu p_F}{\pi^2}\left(2\mu^2+m^2\right)
    %     \nonumber\\ &\quad
    %     +\frac{3m^4}{4\pi^2}\left[3-4\ln\left(\frac{\mu+p_F}{\Lambda}\right)\right]
    &\nonumber\\
    F^{(0)}_\text{\tiny pQCD}[\Pi_5]    &= \frac{6m^2}{\pi^2}\mu p_F
            \nonumber\\ &\!\!\!
        +\frac{3m^4}{\pi^2}\left[1-2\ln\left(\frac{\mu+p_F}{\nu}\right)\right],\\
        &\nonumber\\
    F^{(0)}_\text{\tiny pQCD}[\Psi_5]    &= \frac{6m^2}{\pi^2}\mu p_F(2\mu^2-m^2)
            \nonumber\\ &\!\!\!
    +\frac{3m^6}{2\pi^2}\left[3-4\ln\left(\frac{\mu+p_F}{\nu}\right)
    \right],
\end{align}
 with $p_F=\sqrt{\mu^2-m^2}$.
% In the limit $\mu=\mu_c$, the Fermi momentum turns out to be $p_F\to \sigma\sqrt{s_0}/2$, and the FESR at $\mu>\mu_c$ coincides with the FESR at $\mu<\mu_c$, except Eq.\,\eqref{eq:FpQCD^0[Piss]-mu>muc}. 
% Precisely $F^{(0)}_\text{\tiny pQCD}[\Pi_0]$ contains the scattering contribution. 
% In the case of $F^{(0)}_\text{\tiny pQCD}[\Pi_0^{ss}]$, the transition is discontinuous. 
% If the scattering contribution be absent, then there remains an explicit $\mu$ dependence in the FESR at $\mu<\mu_c$.

\subsection{Quark condensates}
The Wilson coefficients for the quark condensates are tree-level. 
The non-vanishing contribution to the FESR in terms of $\langle \bar qq\rangle$ and $\langle \theta_q\rangle$ are already displayed in Eqs.\,\eqref{eq:Fqq-1-ts}-\eqref{eq:Fqq-1-s} for odd $n$, and Eqs.\,\eqref{eq:FESR_mu_Pi0+0}-\eqref{eq:Pi5-} for even $n$,  where $n$ is the power in weight function $\omega^{n+1}$. 

We will now present the contribution of quark condensates as they appear in their ``raw" form, without any further manipulations, such as applying the quark equations of motion.
Since quark condensates produces an operator mixing, it is customary to write the FESR keeping explicit the dimension $d$, because it affects the value of the operator mixing described in Eq.\,\eqref{eq:a_mu-nu}.
The non-vanishing condensates for even $n$ are
\begin{align}
    F^{(2)}_{qq}[\Pi^{ts}_0]    &=  -2\sum_{q=u,d}\left[\langle \bar q\gamma_0 iD_0 q\rangle +\frac{\langle \bar q \gamma^j iD_j q\rangle}{d-1} \right], \\
    F^{(2)}_{qq}[\Pi^{ss}_0]    &=  -\frac{4}{d-1}\sum_{q=u,d}\langle \bar q \gamma^j iD_j q\rangle,
    \\ &\nonumber\\
    F^{(0)}_{qq}[\Psi_5]        &=  -8m^2\sum_{q=u,d}\langle \bar q\gamma_0 iD_0 q\rangle,
\end{align}
recalling the relation $\langle \bar q i\slashed D q\rangle = m_q\langle \bar q  q\rangle$ with $q=u,d$.

In the case of FESR with odd $n$  in the weight function, there is no operator mixing since the quark number operator appears as the difference of $u$ and $d$ number densities, and in the equal quark mass approximation, the mixing vanishes. 
%If different quark masses are considered, the operator mixing depends exclusively on quark chemical potential. 
It is worth to stress that in the calculation of $F_{qq}^{(1)}[\Pi_5]$, the following relation was considered
\begin{equation}
    \langle q^\dag \gamma_\mu iD_\nu q\rangle = u_\mu u_\nu m_q\langle q^\dag q\rangle, 
\end{equation}
using equations of motion and translation invariance \cite{Jin:1992id}.

\subsection{Gluon condensates}

    The gluon contribution to the FESR is split in terms of gluon condensate and gluon energy density contributions, namely,
\begin{equation}
        F^{(n)}_{GG}[\Pi] = A^{(n)}_{\Pi}\langle G^2\rangle + B^{(n)}_{\Pi}\langle \theta_g\rangle.
        \label{eq:FESR_gluon-notation}
\end{equation}

Then, the gluon condensate contribution is
\begin{align}
    A^{(0)}_{\Pi_0} &= \frac{m^4}{3s_A^3}\left[\frac{1}{\sigma_A^3}\right],\\
    A^{(2)}_{\Pi_0} &= -\frac{1}{12} \left[\frac{1}{4\sigma_A^3}\left(-1+8\sigma_A^3-3\sigma_A^4\right)\right] , \\
    A^{(0)}_{\Pi_5} &= \frac{m^4}{s_P^2}\left[\frac{1}{3\sigma_P^3}\left(1+\frac{8 \sigma_P^2}{ (1 + \sigma_P)^2}\right)\right],\\
    A^{(0)}_{\Psi_5}&= \frac{m^2}{2}\left[\frac{1}{6\sigma_P^2}\left(1+9\sigma_P-4\sigma_P^2\right)\right],
\end{align}
and the gluon-energy density condensate contribution is
\begin{align}
    B^{(0)}_{\Pi_0^{ts}}  &=  \frac{4}{9s_A}\left[\frac{1}{16\sigma_A^3}\left(-3 +6\sigma_A^2+13\sigma_A^4\right) \right],\\
      &\nonumber\\
    B^{(0)}_{\Pi_0^{ss}}  &=  -\frac{4}{9s_A}\left[ \frac{1}{16\sigma_A^3}\left( 3+30\sigma_A^2-17\sigma_A^4\right) \right],\\
     &\nonumber\\
% \end{align}
% \begin{align}
    B^{(2)}_{\Pi_0^{ts}}  &=  \frac{23}{27}\left[\frac{1}{92\sigma_A^3}\left(-9-16\sigma_A^3+117\sigma_A^4\right)\right]
        \nonumber\\ & \quad
                                -\frac{8}{9}\ln \left(\frac{\sqrt{s_A}}{\nu}\,\frac{1+\sigma_A}{2}\right),\\
     &\nonumber\\
    B^{(2)}_{\Pi_0^{ss}}  &=  \frac{1}{27}\left[\frac{1}{4\sigma_A^3}\left(-9-108\sigma_A^2-32\sigma_A^3+153\sigma_A^4\right)\right]
                             \nonumber\\ & \quad  
                                +\frac{8}{9}\ln \left(\frac{\sqrt{s_A}}{\nu}\,\frac{1+\sigma_A}{2}\right),
      \\&\nonumber\\
% \end{align}%
% \begin{align}
    B^{(0)}_{\Pi_5^t}   &=  \frac{8m^2}{3s_P}\left[ \frac{1}{8\sigma_P^3}\left(-1+9\sigma_P^2\right)\right],\\
     &\nonumber\\
    B^{(0)}_{\Pi_5^s}   &=  -\frac{8m^2}{3s_P}\left[ \frac{1}{8\sigma_P^3}\left(1+7\sigma_P^2\right)\right],\\
    &\nonumber\\
    B^{(0)}_{\Psi_5}    &=  \frac{2m^2}{3}\left[ \frac{1}{2\sigma_P^3}\left( -1+7\sigma_P^2-4\sigma_P^3\right)\right] 
                    \nonumber\\ & \quad   -\frac{16m^2}{3}\ln \left(\frac{\sqrt{s_P}}{\nu}\,\frac{1+\sigma_P}{2}\right).
\end{align}

In all gluon-related Wilson coefficients, the content inside the squared parentheses gives $[\cdots]_{\sigma\to 1}=1$ which corresponds to the limit $m^2/s_0\to 0$.
Notice that for gluon condensate $\langle G^2\rangle $, there is no difference between temporal and spatial components in $\Pi_0$ and $\Pi_5$, so we suppress temporal and spatial indexes.
The explicit dependence on the subtraction scale appears only in the contributions with  $\langle \theta_g\rangle$.

\end{appendices}

% \bibliography{FESR_mu}

\end{document}